# Disjoining Pressure Driven Transpiration of Water in a Simulated Tree

Sajag Poudel, An Zou, Shalabh C. Maroo*

Department of Mechanical & Aerospace Engineering, Syracuse University, NY 13244

*scmaroo@syr.edu

**Abstract:** We present an investigation of transpiration of water in a 100 m tall tree using continuum simulations. Disjoining pressure is found to induce absolute negative pressures as high as -23.5 atm at the liquid-vapor meniscus during evaporation, thus presenting a sufficient stand-alone explanation of the transpiration mechanism. In this work, we begin by first developing an expression of disjoining pressure in a water film as a function of distance from the surface from prior experimental findings. The expression is then implemented in a commercial computational fluid dynamics solver and the disjoining pressure effect on water wicking in nanochannels of height varying from 59 nm to 1 micron is simulated. The simulation results are in excellent agreement with experimental data, thus demonstrating and validating that near-surface molecular interactions can be integrated in continuum numerical simulations through the disjoining pressure term. Following the implementation, we simulate the transpiration process of passive water transport over a height of 100 m by using a domain comprising of nanopore connected to a tube with a ground-based water tank, thus mimicking the stomata-xylem-soil pathway in trees. By varying the evaporation rate from liquid-vapor interface in the nanopore, effects on naturally-created pressure difference and liquid flow velocity are estimated. Further, kinetic theory analysis is performed to study the combined effect of absolute negative liquid-film pressure and accommodation coefficient on the maximum mass flux feasible during transpiration. Continuum simulations coupled with kinetic theory reiterate the existence of an upper limit to height of trees. The numerical model developed here is adept to be employed to design and advance several other nanofluidics-based natural and engineering systems.

## 1. Introduction

Stomatal transpiration in trees is theorized to generate the required pressure difference to passively drive water from ground to the leaves.[1,2] Trees self-regulate the size of stomata pore opening which can range between micrometers[3] to a few nanometers[4]. In tall trees such as redwoods[5] which are 100 m high, a pressure difference of more than 10 times the atmospheric pressure is required to pull water against gravity from the soil to the leaves. Thus, assuming water in the soil is at 1 atm, absolute negative pressure of more than -11 atm is supposed to occur in stomata pores. The overall process involves the evaporation of water at liquid vapor interface in stomatal opening potentially generating such large pressure difference enabling the passive flow of water. Thus, comprehension of the absolute negative pressure in liquid and the passive propagation of water in nanoscale pores is crucial to build the fundamental understanding of the process.

The absolute pressure in a nanoscale thin liquid film is significantly reduced from bulk due to solid-liquid interatomic interactions; such a reduction is characterized by the well-established disjoining pressure theory[6]. The effect of disjoining pressure on liquid film pressure is mathematically captured through the modified Young-Laplace equation.[7] Disjoining pressure also occurs in several other natural and engineering processes where thin liquid films are ubiquitous, such as in heat-transfer,[7-18] droplet-spreading,[10,11,19] and those involving bubbles,[7,20-22] etc. The quantitative contribution of interatomic interaction to disjoining pressure is dependent on the atomic composition and distance from the surface. It is mainly attributed to van der Waal's force if either solid or liquid atoms are non-polar; long-range electrostatic forces dominate when both solid and liquid atoms are polar. Since many natural phenomena and engineering applications involve the polar fluid water, we focus on the scenario comprising water – silicon dioxide combination where electrostatic forces can affect molecular motion of water up to tens to hundreds of nanometers from the surface[23-25]. On the basis of the quantification of disjoining pressure for polar liquid in a Gibbsian composite system[26,27], here we develop a numerical understanding of transport of water in nanoscale pores and its potential impact on transpiration in 100 m tall redwood trees.

In current literature, theoretical estimation of disjoining pressure of polar molecular combinations, i.e. water – silicon dioxide, using DLVO theory requires approximation of surface potential;[28] while discrete numerical simulations, such as molecular dynamics, are currently limited by the computational ability to simulate large domains. In our recent work,[27] the disjoining pressure of water was quantitatively characterized by conducting wicking experiments in 1-D silicon dioxide nanochannels. Disjoining pressure was found to dominate the driving mechanism for the liquid wicking in low height nanochannels ($h$ < 100 nm). The average value of disjoining pressure ($\widehat{P_d}$) in the water film was deduced as an exponential function of the film thickness ($\delta$) (i.e. half of the channel height);[27] however, due to the challenges of measuring local pressure at locations away from the surface at the nanometer scale, the knowledge of the disjoining pressure distribution in water film as a function of the distance ($y$) from the surface, $P_d(y)$, is still lacking[6,29]. In this work, we first derive an expression of $P_d(y)$ based on our prior experimental findings, then integrate $P_d(y)$ in continuum simulations, and finally apply the developed numerical methodology to investigate the transpiration mechanism. For this purpose, two different computational domains are considered. First, to validate the numerical technique of integrating the effect of disjoining pressure $P_d(y)$ in the Computational Fluid Dynamics (CFD) simulation, wicking in nanochannels of geometry consistent to the experimental study[27] is investigated and presented in Section 2. Second, in Section 3, the developed method is employed in the system consisting of nanopore-tube-ground water tank to resemble the transpiration process in 100 m tall trees.

## 2. Methods

### 2.1 Expression of disjoining pressure

We first develop an expression to estimate disjoining pressure of water as a function of distance from the solid wall $P_d(y)$, following which we implement the disjoining pressure effect in a commercial CFD solver by supplementing an additional source term in the momentum equation. The implementation is then utilized to simulate the wicking process of water in nanochannels connected to a reservoir (Fig. 1a). Our work integrates and validates a molecular level phenomenon (disjoining pressure) into continuum simulations, and can be further used to study nanoscale effects as well as design engineering systems involving nanoscale liquid flows.

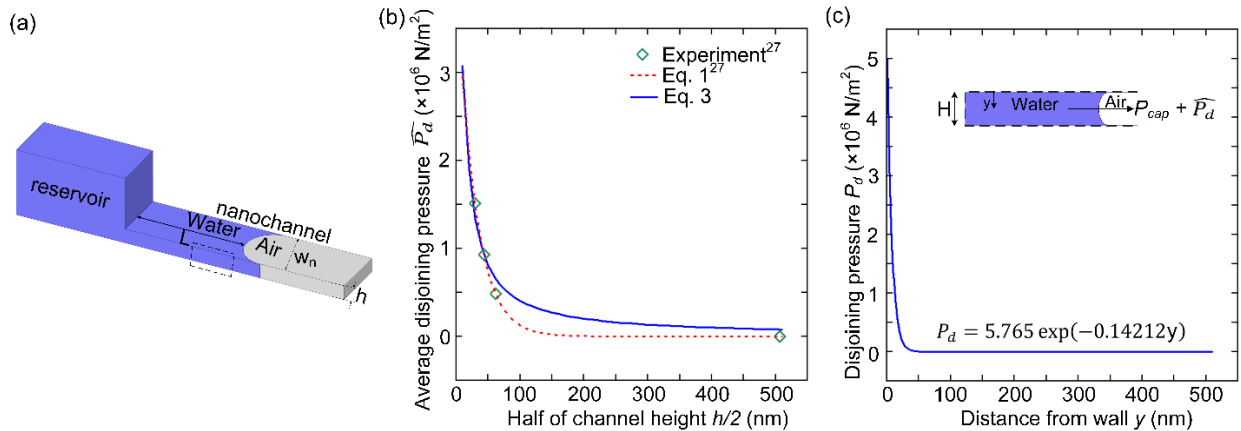

*Figure 1. Wicking in nanochannels. (a) Schematic of water wicking in a nanochannel connected to a reservoir. (b) Variation of average disjoining pressure existing in a nanochannel with channel half-height. (c) Variation of disjoining pressure with distance from the wall.*

Equation 1 represents the average value of disjoining pressure of water in nanochannel ($\widehat{P_d}$) as a function of the liquid film thickness (=$h/2$ which is half the nanochannel height) obtained from curve fit of experimental data in our recent work[27] (Fig. 1b). In order to develop an expression of $P_d$ as a function of the distance from the surface ($y$), we adopt a similar exponential function as shown in Eq. 2, where $A$ and $B$ are constants to be determined. By integrating Eq. 2 with respect to $y$ as shown in Eq. 3, the average disjoining pressure of water in nanochannel can be obtained, and then compared against Eq. 1 to estimate the constants as explained next.

$$\widehat{P_d} = 4.25 \exp\left(-0.035 \frac{h}{2}\right) \quad \text{Equation 1}$$

$$P_d = A \exp(-By) \quad \text{Equation 2}$$

$$\widehat{P_d} = \frac{1}{h}\int_0^h P_d(y)\,dy = \frac{1}{h}\int_0^h A\exp(-By)\,dy = \frac{A[1-\exp(-Bh)]}{hB} \quad \text{Equation 3}$$

Constants $A$ and $B$ were determined by an iterative process. First, random values were assigned in Eq. 3, whose results were compared to that from Eq. 1. The values were converged in each iteration by minimizing the error between $\widehat{P_d}$ computed from Eq. 1 and Eq. 3. The convergence is obtained by estimating the coefficient of determination $R^2$ between the solutions of Eqs. 1 and 3 for '$N$' number of different nanochannel height cases with the corresponding $h$. Based on this analysis, the best fits of $A$ and $B$ are selected for the highest and the most stable value of $R^2$ (i.e., the value of $R^2$ which would be independent of the number of data-points considered). Eventually, constants $A$ and $B$ were finalized as 5.765 and 0.14212 respectively (see Eq. 4), with $R^2$ = 91.3% resolved to the precision of $\Delta y$ = 1 nm with ~500 data points. Additional details on the iterative process to obtain values of $A$ and $B$ are also provided in supplementary section S1.

$$P_d = 5.765 \exp(-0.14212y) \quad \text{Equation 4}$$

Substituting these values of A and B, the averaged disjoining pressure ($\widehat{P_d}$) from Eq. 3 is in good agreement with both experimental data and fitting curve (Eq. 1) prediction (Fig. 1b). The corresponding disjoining pressure distribution $P_d$ in water film is plotted in Fig. 1c.

*Table 1. Nanochannels geometry and corresponding contact angles at top and side walls[27]*

| Case | Reservoir (Microchannel Geometry) | | Nanochannel Geometry | | Contact Angle on Nanochannel walls | |
|---|---|---|---|---|---|---|
| | Height ($h_m$) | | Channel height ($h$) | | Top Wall | Side Wall |
| 1 | 1.7 µm | Length ($L_m$) = 150 µm, Width ($w_m$) = 40 µm | 59 nm | Channel Length ($L_n$) = 150 µm, Channel width ($w_n$) = 10 µm | 90° | 29.4° |
| 2 | 2.5 µm | | 87 nm | | 90° | 27.3° |
| 3 | 3.5 µm | | 124 nm | | 64.8° | 40.6° |
| 4 | 30 µm | | 1015 nm | | 39.6° | 39.6° |

## 2.2 Integration to CFD simulation

Next, to simulate the wicking process in nanochannels of geometry consistent with experiments[27] (Table 1), disjoining pressure effect is implemented in continuum simulation using a laminar multiphase model with the volume of fluids method. A commercial CFD software ANSYS Fluent is utilized to solve the governing equation of fluid flow and a user defined function (*udf*) is linked to the momentum equation to specify the source term based on the disjoining pressure function.[30,31] As shown in Fig. 2a, a nanochannel of height ($h$), length ($L_n$), and width ($w_n$) is connected to the reservoir of height ($h_m$), length ($L_m$), and width ($w_m$). $L_m$ and $w_m$ were fixed as 150 µm and 40 µm respectively, while $h_m$ changed with nanochannel height ($h$) with a fixed ratio ($h_m/h$) of ~30. Four channel heights ($h$) of 59 nm, 87 nm, 124 nm, and 1015 nm were simulated with fixed channel length $L_n$ of 150 μm, and width $w_n$ of 10 μm (Table 1). Based on the advantage of symmetry, only one-fourth portion of the physical domain is considered for the CFD simulation (Fig. 2b). The faces AB and EF towards the reservoir and the nanochannel respectively are open with the pressure outlet (1 atmosphere) boundary condition. The face ADF is symmetric and no-slip condition is set at all walls of nanochannel and reservoir. Surface tension is evoked using a continuous shear force model[32] (liquid-air surface tension $\gamma$ = 0.072 N/m) together with wall adhesion and specified contact angles. The contact angles on the side and top walls of each case of nanochannel height are adopted from our prior experimental work[27] and are listed in Table 1. While the values of the contact angles for the small and large height nanochannels presented in Table 1 are obtained directly through an analysis based on experimental visualization and molecular dynamic simulation[27], the contact angle on top wall of the nanochannel with intermediate height $h$ = 124 nm is not explicitly available. Thus, the same is deduced here employing an analysis based on the average disjoining pressure. In the earlier experimental study[27], the average disjoining pressure in the case of nanochannel with $h$ = 124 nm is considered as a mean of two different scenarios (denoted as $\widehat{P_{d-m}}$) corresponding to the contact angles on the top wall ($\theta_{t-w}$) of the channel: 90° and 39.6° respectively, while the contact angle on the side wall being ($\theta_{s-w}$ = 39.6°). For the CFD simulation, we introduce a new contact angle on the top wall ($\theta_{t-w}^*$), which would induce a disjoining pressure equivalent to $\widehat{P_{d-m}}$.

From the previous work[27], $\widehat{P_{d-m}}$ = 452,488.6 $Pa$

Hence,

$$\widehat{P_{d-m}} = \frac{6\mu C^2}{h^2} - 2\sigma\left(\frac{\cos\theta_{t-w}^*}{h} + \frac{\cos\theta_{s-w}}{w}\right) \qquad \text{Equation 5}$$

Where, $C$ is the slope of fitted curve in wicking distance v/s $t^{1/2}$ plot; $C$ = 1.62 for nanochannel with $h$ = 124 nm, $w$ = 10 μm[27].

Solving Eq. 5, we get $\theta_{t-w}^*$ = 64.8° (also see Table 1) which is utilized for simulation in case of $h$ = 124 nm.

Figure 2c illustrates a magnified view of the XY plane of the domain with the plane of symmetry (green line) and Fig. 2d demonstrates mesh refinement near the top wall of the nanochannel.

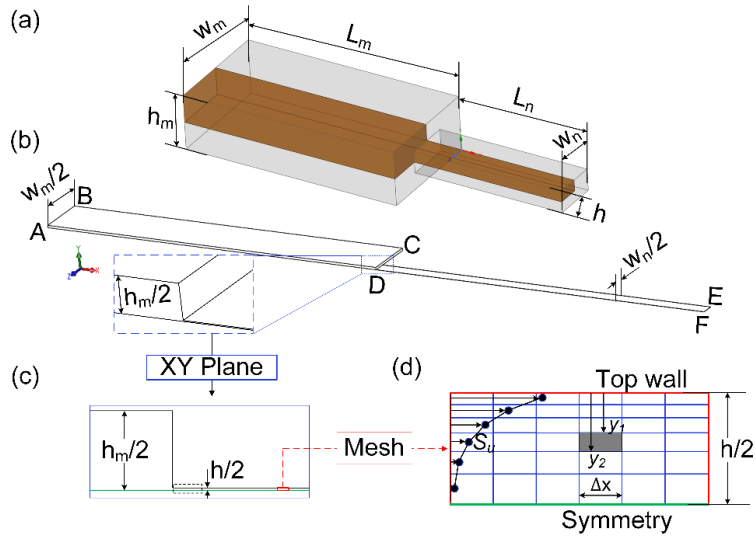

Figure 2. CFD Simulation of water wicking in a nanochannel. (a) Sketch of a nanochannel connected to a reservoir showing symmetry. (b) Computational domain utilized in the present study. (c) XY plane view of the domain showing the plane of symmetry. (d) Mesh refinement near the wall of nanochannel together with the specified source term profile along the height.

The effect of the disjoining pressure (Eq. 4) is included by adding a source term $S_u$ in the liquid phase, which is added to the discretized form of the X-momentum equation through a *udf* as follows[30]:

$$S_u = \int_{y_1}^{y_2} \frac{P_d}{\Delta x} \qquad \text{Equation 6}$$

where, $y_1$ and $y_2$ are limits of the finite volume cell in Y-axis and $\Delta x$ is the cell width along X-axis (see Fig. 2d). The stated source term in the corresponding momentum equation is estimated at the center of each finite volume cell and acts as a supplementary driving force for liquid propagation. Moreover, the computational domain is discretized non-uniformly with the refined grid spacing (smaller finite volume cells) near the wall (see Fig. 2d) to better capture the exponential effect of disjoining pressure. Additional details on wicking simulation with uniform and non-uniform grids along with the grid-independence test and the Richardson's error estimation[33] for relative error in simulation is provided in supplementary section S2.

**2.3 Wicking in nanochannel**

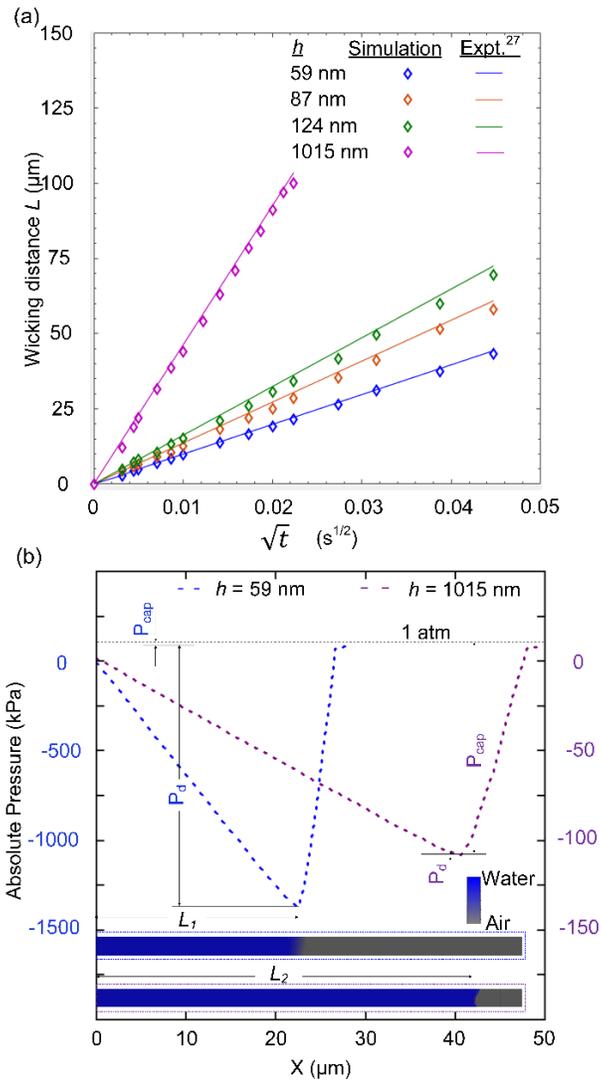

*Figure 3. (a) Evolution of wicking distance with time for nanochannels of varying height. (b) Absolute pressure along the nanochannel length indicating the peak negative pressure at the meniscus. The inset (corresponding to the dashed-box of Fig. 2c) shows the liquid phase volume fraction contour plot of the nanochannel.*

Figure 3a shows the evolution of wicking distance *L* with time from CFD simulations for all four channel heights. By including the effects of both $P_{cap}$ and $P_d$, the simulation data are in good agreement with experimental data.[27] Thus, the expression of disjoining pressure $P_d(y)$, as a function of distance from the surface $y$ (Eq. 4), for water – silicon dioxide combination, as well as the implementation of disjoining pressure effect to CFD solver is demonstrated to be accurate as they adequately reproduce the experimental outcome. Further information and numerical values of $S_u$ for each case of channel height are also provided in supplementary Section S3. The findings from the present work can be used to further study phenomenon where near-surface effects on liquid behavior are non-trivial.

The variation of pressure along the direction of wicking in 59 nm and 1015 nm channels is shown in Fig. 3b. During this process, there exists a pressure gradient along the direction of wicking with a peak negative pressure ($P_n$) at the meniscus,[11,13,34] and is estimated from the corresponding contour plots (insets of Fig. 3b) of liquid volume fraction inside the channel. The capillary pressure is obtained by $P_{cap} = 2\sigma \left( \frac{\cos \theta_{side}}{h} + \frac{\cos \theta_{top}}{w} \right)$, with contact angle values listed in Table 1; while the value of average disjoining pressure is obtained from Eq. 1. Table 2 lists all values of $P_{cap}$, $\widehat{P_d}$ and $P_n$ for all channel heights. For small channel heights (<100 nm), $P_d$ dominates and thus the relation $P_d \approx P_a - P_n$ holds; while in 1015 nm channel, $P_d$ is negligible and thus the relation $P_{cap} \approx P_a - P_n$ applies. From this investigation of liquid film pressure, the relative contribution of $P_{cap}$ and $P_d$ on the overall driving mechanism is found to widely vary for the two extreme heights of nanochannels considered in the present study.

*Table 2. The values of different pressures associated with each case of nanochannel height.*

| Channel height (h) | $P_{cap}$ | $\widehat{P_d}$ | $P_n$ |
|---|---|---|---|
| 59.6 nm | 1.25×10⁴ | 1.39×10⁶ | -1.40×10⁶ |
| 87 nm | 1.28×10⁴ | 9.30×10⁵ | -8.78×10⁵ |
| 124 nm | 4.46×10⁵ | 6.54×10⁵ | -8.23×10⁵ |
| 1015 nm | 1.09×10⁵ | 7.91×10⁴ | -1.05×10⁴ |

## 3. Transpiration in Trees

### 3.1 Computational Domain

Following the implementation of the effect of disjoining pressure in CFD simulation, we investigate the liquid transport through xylem in trees driven by evaporation at stomatal pore. Xylem tube of cross section 100 μm × 100 μm is connected to tank at bottom, and on top of the xylem, a nanopore is provided which resembles the stomatal opening in trees. The nanopore is 10 μm in height with a cross-section 59 nm × 10 μm (consistent to the nanochannel[27] and case 1 of Table 1 discussed in Section 2). The nanopore opening is opted to have a rectangular cross section over a circular/elliptical one since the wettability of water in confinement are well established for a flat surface[27] (also see Table 1). In the numerical simulation of transpiration in trees discussed in this section, evaporation (liquid to vapor mass transfer) at the meniscus of nanopore is evoked to represent the stomatal transpiration. Thus, it establishes a numerical system of transpiration driven passive flow of water through xylem in 100 m tall trees.

In the present numerical model, due to the constraint of aspect ratio, height of xylem tube is only 1 cm. However, the operating condition is tuned to resemble a 100 m tall xylem tube by applying gravity equivalent to $10^4 \times g$ (g = 9.81 m/s²) in only the xylem region of the computational domain. All other regions (tank and nanopore) are set with g = 9.8 m/s². As shown in Fig. 4a-b, only one-fourth part of the domain is considered for the simulation based on the advantage of symmetry. Fig 4b also illustrates the geometrical parameters and boundary conditions at different faces of the computational domain. Moreover, the temperature throughout the domain is constant 300 K and the pressure outlet at open boundaries is set at 1 atm. A specified evaporation rate flux (liquid to vapor phase transition) at the meniscus ($\dot{m}_e''$) is induced through a *udf* with mass source term. The effect of varying $\dot{m}_e''$ on the transpiration driven flow is investigated with the focus on the equilibrium meniscus level in nanopore ($h_{np}$), liquid transport velocity in xylem ($u_{xl}$) and in nanopore ($u_{np}$), and the effective pressure difference driving the passive flow ($\Delta P$).

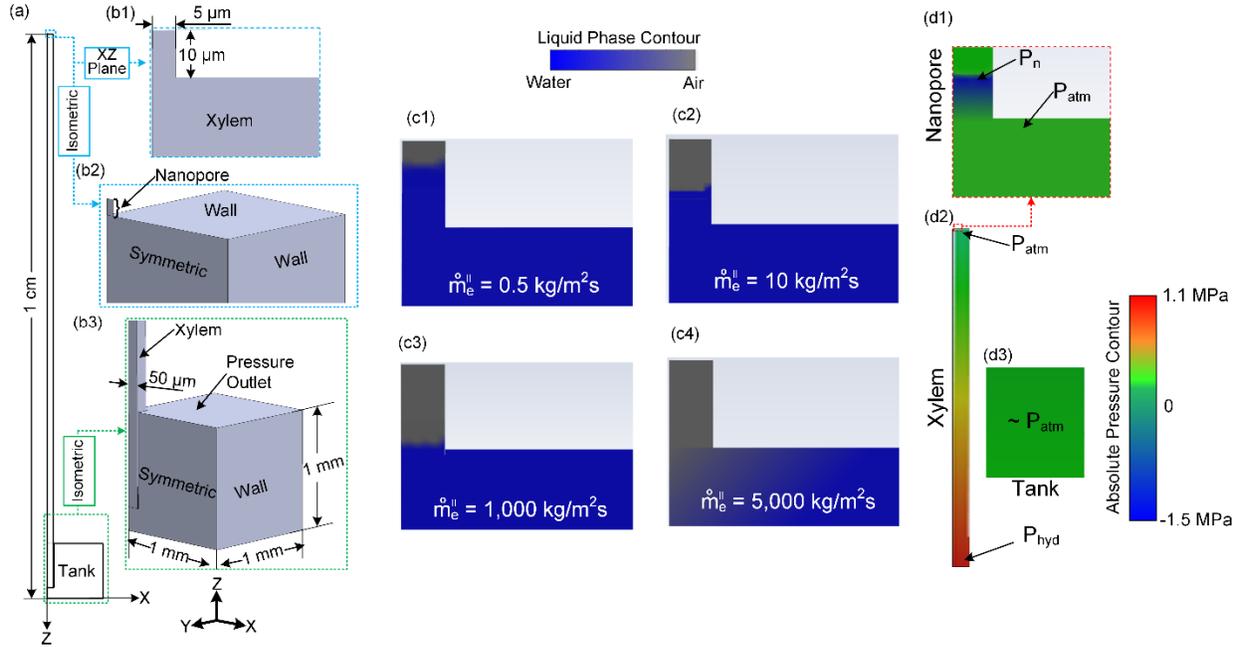

*Figure 4. Numerical simulation of stomatal transpiration and passive water flow in xylem tubes. (a) Computational domain. (b) Different zones of computational domain indicating the boundary conditions. (c) Liquid phase contour plot illustrating the equilibrium position of meniscus in nanopore for different cases of $\dot{m}_e''$. (d) Absolute pressure contour plot showing the variation of disjoining pressure in nanopore and hydrostatic pressure in xylem. The section of xylem shown in d-2 is not to scale.*

### 3.2 Evaporation in nanopore and passive water transport

Based on Fick's law of diffusion, the evaporation rate flux from a typical water meniscus is found to be ~0.5 kg/m²s corresponding to the ambient temperature (T = 300 K) and relative humidity (RH = 30%). With the possible rise in temperature and lowering of ambient humidity, $\dot{m}_e''$ can dramatically increase. Thus, eight different cases are simulated with $\dot{m}_e''$ = 0.5, 2.5, 10, 50, 100, 500, 1000, 5000 kg/m²s. The liquid phase contour plot of the nanopore region indicating the meniscus in equilibrium position for $\dot{m}_e''$ = 0.5, 10, 1000, 5000 kg/m²s is shown in Fig. 4c. For the case with the largest $\dot{m}_e''$ (= 5000 kg/m²s), the evaporation mass flux surpasses the wicking flux causing the meniscus to recede into the xylem tube. This indicates non-physical behavior arising due to unnatural high $\dot{m}_e''$. Thus, the case with $\dot{m}_e''$ = 5000 kg/m²s is excluded from further discussion here onwards. Similarly, Fig. 4d shows the variation of absolute pressure over the domain (nanopore, xylem, and tank) when the meniscus reaches an equilibrium position for the case with $\dot{m}_e''$ = 2.5 kg/m²s. The two key aspects of absolute pressure in the system: disjoining pressure and hydrostatic pressure are revealed from the plot. The peak negative pressure ($P_n$) exists at the meniscus in nanopore while the hydrostatic pressure ($P_{hyd}$) of water in xylem results into the maximum positive pressure at the bottom of xylem.

As the evaporation rate is specified in the simulations, a balance between wicking flux and evaporation mass flux occurs in time and the liquid-vapor interface (i.e., meniscus) reaches an equilibrium stable position in the nanopore. In such an equilibrium, the continuous pull of water from the tank, i.e., against the gravitational pull of 100 m height, takes place in order to replenish the evaporated mass. Accordingly, the equilibrium position of meniscus lies at different heights in the nanopore depending on the rate of evaporation specified. The inset in Fig. 5 shows the sketch of evaporating meniscus in nanopore and liquid transport from ground (tank) where $h_{np}$ is defined as distance of meniscus, at equilibrium, from xylem tube. Figure 5 shows that a significant drop in $h_{np}$ occurs with higher $\dot{m}_e''$ values. The water transport velocity at xylem ($u_{xl}$) and in nanopore ($u_{np}$) is also obtained at equilibrium and plotted in Fig. 5. The transport velocity in xylem is found to be in the order of $10^{-8}$ to $10^{-5}$ m/s for the range of $\dot{m}_e''$ considered. Further, the speed of water transport in nanopore is found to be orders of magnitudes higher than that in xylem due to much smaller cross-sectional area. The numerical values of $u_{xl}$ obtained here in a single xylem tube are lower than the reported rate of water transport in trees (~$10^{-3} - 10^{-4}$ m/s) since the water transport in xylem is driven through evaporation in a single nanopore in our study, while multiple stomata are connected to a single xylem in nature.[35,36]

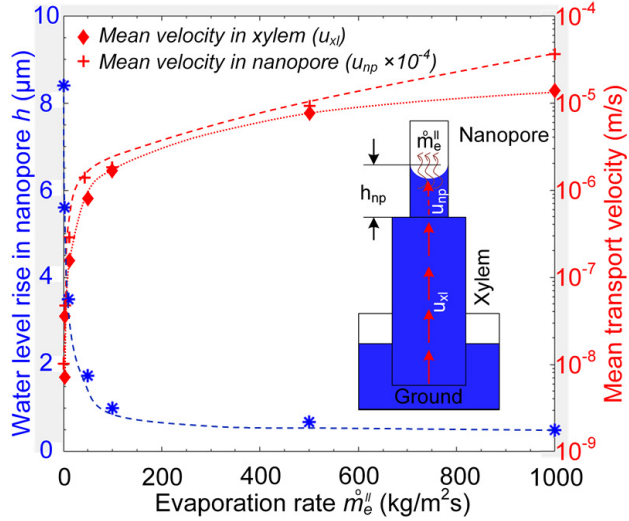

*Figure 5. Variation of equilibrium stage water level in nanopore and mean transport velocity of water with the evaporation rate flux in meniscus. Dotted lines are guides to the eye.*

### 3.3 Pressure difference driving the flow

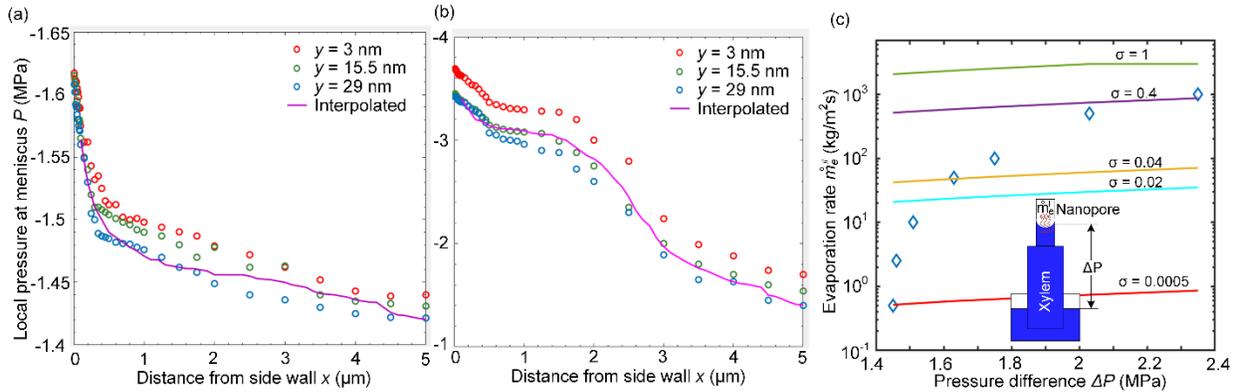

*Figure 6. Variation of local pressure in the meniscus along X-axis at different depth in Y-axis at the nanopore during equilibrium for $\dot{m}_e''$ = (a) 2.5, (b) 1000 kg/m²s. (c) Relationship between the mass transfer rate at liquid vapor interface and the pressure difference driving the passive flow.*

Next, we investigate the pressure at the meniscus to gain insight into the relationship between the stomatal evaporation and the pressure difference driving the passive flow of water. Evaporation at the interface tends to deform the shape of meniscus and induces a local variation of absolute pressure along the meniscus.

Such variation needs to be considered in order to estimate the average $P_n$ for a given case of $\dot{m}_e^"$. Since the finite volume cells (or grid points) in the computational domain are not uniformly distributed (with refined grid spacing near the solid wall), the simple average of the local pressure values at all grid points along the meniscus is not appropriate to estimate $P_n$. Thus, the non-uniformly distributed data points of local absolute pressure in the meniscus along X-axis for each case of $\dot{m}_e^"$ (see Fig. 6a-b) is utilized to interpolate the local pressure values to a temporarily created uniform grid points along X-axis ($\Delta x$ = 0.1 µm). The variation of the pressure at the interpolated data points is also shown in Fig. 6a-b and $P_n$ is computed as the mean of the interpolated data points of corresponding case. Here, inducing the evaporation at the meniscus in nanopore is found to greatly reduce the pressure in liquid film. As compared to the average peak negative pressure at the meniscus of $P_n$ = -1.40 MPa without evaporation (see Table 2), we obtain $P_n$ to be -1.48 MPa and -2.35 MPa for $\dot{m}_e^"$ = 2.5 and 1000 kg/m²s, respectively, during evaporation. Using this methodology, $P_n$ and thus the pressure difference driving the flow i.e., $\Delta P = P_{atm} - P_n$ are computed for all seven cases of $\dot{m}_e^"$ and the variation of $\Delta P$ with $\dot{m}_e^"$ is shown in Fig. 6c. Thus, a higher evaporation rate at the meniscus leads to a greater pressure difference which helps overcome the pressure drops in the nanopore and xylem tube, but at the cost of decreasing h$_{np}$. Eventually, increasing evaporation rate beyond a certain limit causes the meniscus to recede entirely from the nanopore resulting in the failure of the transpiration mechanism (as seen in our simulation case with $\dot{m}_e^"$ = 5000 kg/m²s).

### 3.4 Kinetic theory and mass transfer across interface

Lastly, we performed an analysis of maximum mass transfer across an interface based on kinetic theory.[37,38] The maximum mass flux across a water-vapor interface is given as[37]:

$$\dot{m}_e^" = \frac{2\sigma}{2-p_v} \sqrt{\frac{M}{2\pi R}} \left( \frac{p_{l,corr}}{\sqrt{T_l}} - \frac{p_v}{\sqrt{T_v}} \right) \qquad \text{Equation 7}$$

where, $\sigma$ is accommodation coefficient, ($M$ = 0.018 kg/mol) is the molar mass of fluid, ($R$ = 8.314 J/K.mol) is ideal gas constant, ($T_v$ = 300 K) and ($p_v$ = 101325 Pa) are saturated vapor temperature and pressure respectively, ($T_l$ = 300 K) is temperature of liquid, $p_{l,corr}$ is the corrected liquid pressure which is a function of $p_n$, surface tension of liquid ($\gamma_{lv}$ = 0.072 N/m), radius of pore ($r_p$ ~ 1 µm) and density of liquid ($\rho$ = 1000 kg/m³)[38]:

$$p_{l,corr} = p_n \exp\left(-\frac{2\gamma_{lv}M}{r_p R \rho}\right) \qquad \text{Equation 8}$$

Here the value of $\dot{m}_e^"$ is dependent on accommodation coefficient which varies significantly with different fluids and is sensitive to the molecular properties of fluid.[39] Therefore, $\dot{m}_e^"$ is estimated from Equation 8 over the entire range of $P_n$ obtained in our simulations and for varying $\sigma$ = 0.02, 0.04, 0.4, and 1. As shown in Fig. 6c, the variation of $\dot{m}_e^"$ with $\Delta P$ is found to be strongly dependent on $\sigma$. Furthermore, an additional curve for $\sigma$ = 0.0005 is also created as shown in Fig. 6c which agrees with the lower limit of $\dot{m}_e^"$ in the present work. Figure 6c illustrates that, in order to have a very high passive flow rate of water through xylem, the mass transfer (evaporation) at the meniscus in stomata should be high which in turn necessitates a high value of $\sigma$. Hence, this kinetic theory analysis coupled with our simulation results highlight the importance of three factors: 1) disjoining pressure, 2) evaporation rate, and 3) accommodation coefficient to naturally sustain the driving force to pull water against gravity. A practical limit to these factors, mainly tied to nanopore surface characteristics and water properties, will result in a height limit to how tall trees can grow.[40,41]

### 4. Conclusion

In summary, we utilize the experimental findings of average disjoining pressure of water in nanochannels and derive an expression for disjoining pressure distribution in a water film as a function of distance from the surface. The developed disjoining pressure expression is implemented in computational fluid dynamics solver to perform simulations of wicking in nanochannels and capture near-surface molecular effects. The simulation results are in excellent agreement with the experimental data, demonstrating the successful implementation and validation of disjoining pressure in continuum simulations. The disjoining pressure in the liquid film was found to be the primary driving mechanism for the flow in geometry of characteristic dimension < 100 nm. The implementation of disjoining pressure is further utilized to numerically simulate the transpiration process in trees. The relationship between the evaporation rate at liquid-vapor interface in nanopore and the consequent pressure difference created to drive the passive flow of water through xylem is studied in detail. Evaporation rate at the nanopore meniscus can induce large absolute negative pressures of -2.35 MPa which drive the flow of water through xylem tubes in 100 m tall trees. However, for very high evaporation rates (5000 kg/m$^2$s in our study), the disjoining pressure driven supply of water cannot balance the evaporation at the meniscus resulting in complete dewetting of meniscus from the nanopore and thus the failure of self-driven transpiration mechanism in trees. The numerical investigation is supplemented with an analysis based on kinetic theory which dictates a limit on the maximum mass transfer at the liquid-vapor interface demonstrating the upper limit on height of trees. The present work provides insights to the mechanism of transpiration and passive water transport on trees as well as lays a foundation for future continuum studies of physical phenomenon where nanoscale liquid films are prominent.

**Acknowledgment:** This material is based upon work supported by, or in part by, the Office of Naval Research under contract/grant no. N000141812357. The numerical simulations were performed in the UberCloud virtual machines, an official cloud hosting partner of ANSYS.

**Data Availability Statement:** The data that support the findings of this study are available from the corresponding author upon reasonable request.